\documentstyle[prl,aps,multicol,psfig]{revtex}

\begin{document}
\draft
\title{Mutual Information of Population Codes and Distance Measures in
Probability Space}
\author{K. Kang and H. Sompolinsky}
\address{
Racah Institute of Physics and Center for Neural Computation,\\
Hebrew University, Jerusalem 91904, Israel
}
\date{\today }
\maketitle
\begin{abstract}
We studied the mutual information between a stimulus and a system consisting
of stochastic, statistically independent elements that respond to a
stimulus. Using statistical mechanical methods the properties of the Mutual
Information (MI) in the limit of a large system size, N, are calculated. For
continuous valued stimuli, the MI increases logarithmically with N and is
related to the log of the Fisher Information of the system. For discrete
stimuli the MI saturates exponentially with N. We find that the exponent of
saturation of the MI is the Chernoff Distance between response probabilities
that are induced by different stimuli.
\end{abstract}

\address{\\
Racah Institute of Physics and Center for Neural Computation,\\
Hebrew University, Jerusalem 91904, Israel\\
}

\begin{multicols}{2}
Population coding serves as a common paradigm of information processing in
the brain. Its starting point is the fact that very often the response of a
single neuron to a stimulus is noisy and is only weakly tuned to changes in
the stimulus value. Hence, the information carried by a typical neuron is
rather low. The brain may overcome this limitation by distributing the
information across a large number of neurons which together carry accurate
information about the stimulus\cite{rieke}. This paradigm has inspired
numerous studies of the statistical efficiency of population codes \cite%
{paradiso}-\cite{brunel}. Of particular interest is the way
in which the amount of information about the stimulus depends on both 
the size of the responding population and the response properties 
of the individual neurons.

Many theoretical studies\cite{seung,rissanen,brunel} have employed the
well-known concept of Fisher Information (FI)\cite{cover}. The FI is related
to the derivative of the population response probability with respect to the
stimulus. For a statistically independent population, the FI is an extensive
quantity and is relatively easy to calculate. However, it is restricted to
the case of a continuously varying stimulus. An alternative measure,
applicable for arbitrary stimulus spaces, is provided by Shannon Mutual
Information (MI)\cite{cover}. Unfortunately, except for special cases, exact
calculation of the MI for a large population is difficult even for
independent populations . This is because MI is bounded from above by the
stimulus entropy, and thus is not an extensive quantity. Recently, a
relationship between the MI of a continuous stimulus and the FI in a large
population has been derived\cite{rissanen,brunel}. However, little
theoretical progress has been made on the properties of the MI in large
systems with discrete stimuli\cite{samengo}.

In this paper we introduce statistical mechanical methods to calculate
analytically the behavior of the MI as the system size, $N$, of the
population grows. For continuous valued stimuli our theory yields a
logarithmic dependence of the MI on the FI, in agreement with previous results %
\cite{rissanen,brunel}. In the case of discrete stimuli, the MI saturates
exponentially fast with $N$. We show that the exponential saturation
rate is dominated by a contribution from the two stimulus values that
induce the closest response-probabilities in the population. The
contribution of this pair of stimuli to the saturation rate equals the
Chernoff Distance between the corresponding response-probabilities, of Large
Deviation Theory \cite{cover} .

We consider a population of $N$ stochastic units, which we call neurons,
that respond simultaneously to a presentation of a stimulus. We denote their
responses by a vector ${\bf r}=\{r_{1},r_{2},...,r_{N}\}$ where $r_{i}$
represents the stochastic response of the i-th neuron to a stimulus. The
stimulus states are denoted by the scalar variable ${\theta }${, which can be%
} either discrete or continuous, with a prior probability (or density) $%
p(\theta )$. We denote the probability (or the density) of ${\bf r}$ given a
stimulus ${\theta }$ by $P({\bf r}|{\theta }).$ In this paper we will focus
on the case of statistically independent neurons (given a stimulus $\theta $%
), namely, 
\begin{equation}
P({\bf r}|\theta )=\prod_{i=1}^{N}P_{i}(r_{i}|\theta ).  \label{ind}
\end{equation}%
An important issue is how to quantify the efficiency of the coding of $%
\theta $ in the population responses. For probabilities that are
differentiable functions of a continuously varying stimulus, a well-known
measure of the efficiency of the population code is the FI. In our case it
is 
\begin{eqnarray}
J(\theta ) &=&-\sum_{i=1}^{N}\left\langle {\frac{\partial ^{2}}{\partial
\theta ^{2}}}\ln P(r_{i}|\theta )\right\rangle _{r|\theta }  
\label{Fisher2} \nonumber\\
&=&\sum_{i=1}^{N}\left\langle \left\{ {\frac{\partial }{\partial \theta }}%
\ln P(r_{i}|\theta )\right\} ^{2}\right\rangle_{r|\theta}
\end{eqnarray}%
where $\langle \cdots \rangle _{r{\bf |\theta }}$ denotes an average with
respect to $P(r_{i}|\theta )$, which is clearly an extensive quantity. 
Here we study an alternative measure of efficiency, the MI of the system,
which is the average amount of information on ${\theta }$ that is added by
observing the response, ${\bf r}$. It is useful to define a local MI, $%
I(\theta )$, 
\begin{equation}
I(\theta )\equiv \left\langle \log _{2}\left\{ \frac{P({\bf r|\theta })}{P(%
{\bf r})}\right\} \right\rangle _{{\bf r|\theta }}  \label{Itheta}
\end{equation}%
where $P({\bf r})={\sum_{\theta }}P({\bf r|\theta })p(\theta )$, and $%
\langle \cdots \rangle _{{\bf r|\theta }}$ denotes an average w.r.t. $P({\bf %
r}|\theta ).$ The full MI, $I$, is 
\begin{equation}
I=\left\langle I(\theta )\right\rangle _{\theta }  \label{Ifull}
\end{equation}%
where $\langle \cdots \rangle _{\theta }$ is an average over the stimulus
distribution. A central quantity is the log-likelihood variable, 
\begin{equation}
S({\bf r,}\phi ,\theta )=\ln \left[ \frac{P({\bf r|\phi })}{P({\bf r|\theta }%
)}\right] \ =\ \sum_{i=1}^{N}S_{i}(r_{i},\phi ,\theta )  \label{loglike}
\end{equation}%
where $S_{i}(r_{i},\phi ,\theta )=\ln P(r_{i}|\phi )-\ln P(r_{i}|\theta )$.
Equation (\ref{Itheta}) can be written as 
\begin{equation}
I(\theta )=-\int \prod_{\phi \neq \theta }dS_{\phi }P_{\theta }\{S_{\phi
}\}\log _{2}\left\langle \exp S_{\phi^{\prime} }\right\rangle _{\phi^{\prime}}  \label{DH}
\end{equation}%
where $P_{\theta }\{S_{\phi}\}$ is the distribution of $S_{\phi }$ $\equiv
S({\bf r,}\phi ,\theta )$, calculated with respect to $P({\bf r|\theta }).$
For large $N$ this distribution is centered around the mean value of $%
S_{\phi }$ 
\begin{equation}
\sum_{i=1}^{N}\left\langle S_{i}(r_{i},\phi ,\theta )\right\rangle
_{r|\theta }=-D_{KL}(\phi ||\theta )  \label{dkl}
\end{equation}%
where $D_{KL}(\phi ||\theta )>0$ is the relative entropy of $P({\bf r|\phi }%
) $ and $\,P({\bf r|\theta }),$ also known as the Kullback-Leibler (KL)
distance between the two distributions. The correlation matrix of the
fluctuations $\delta S_{\phi }=$ $S_{\phi }+D_{KL}(\phi ||\theta )$ is also
of order $N$, and is given by%
\begin{equation}
C(\phi ,\psi |\theta )=\sum_{i=1}^{N}\left\langle \delta S_{i}(r_{i},\phi
,\theta )\delta S_{i}(r_{i},\psi ,\theta )\right\rangle _{r|\theta }.
\label{Cpw}
\end{equation}

We first discuss the MI of a large system with a continuous varying stimuli.
Here, all averages over stimulus space stand for integrals with a density $%
p(\theta )$, and we assume that the probabilities vary smoothly with $\theta
.$ In this case, the dominant contribution to Eq.~(\ref{DH}) comes from
values of $S_{\phi }$ near their mean value and $\phi $ which is near $%
\theta $, i.e., $|\phi -\theta |\lesssim 1/\sqrt{N}.$ This is because  
$S_{\theta }= S({\bf r,}\theta ,\theta ) =0$ while
for $\phi$ that is far from $\theta$, $S_{\phi }\simeq -D_{KL}(\phi
||\theta )$, which is large and negative. For small magnitudes of $\delta \phi $ =$%
\phi -\theta $, and $\delta \psi $ =$\psi -\theta $ we can write 
\begin{equation}
D_{KL}(\phi ||\theta )\approx {\frac{1}{2}}J(\theta )\delta \phi
^{2},\,\;C(\phi ,\psi |\theta )\approx J(\theta )\delta \phi \delta \psi
\label{DCJ}
\end{equation}%
where $J(\theta )$ is the FI, Eq.~(\ref{Fisher2}). The low rank form of the
correlation matrix in Eq.~(\ref{DCJ}) implies that the fluctuations $\delta
S_{\phi }$ can be described as $z\sqrt{J(\theta )}\delta \phi $, where by
central limit theorem $z$ is a Gaussian variable with zero mean and unit
variance. Substituting this in Eq.~(\ref{DH}) yields, 
\begin{eqnarray}
&&I(\theta )\approx -\int Dz\times  \nonumber \\
&&\log _{2}\left\{ \int d\psi p(\psi +\theta )\exp \left( -\frac{J(\theta
)\psi ^{2}}{2}+z\sqrt{J(\theta )}\psi \right) \right\}  \label{IGauss}
\end{eqnarray}%
where $Dz=dz\exp (-z^{2}/2)/\sqrt{2\pi }.$ Evaluating the integrals in the
limit of a large $J$, and substituting in Eq.~(\ref{Ifull}), yields 
\begin{equation}
I\approx H_{\theta }-\frac{1}{2}\left\langle \log _{2}\left\{ \frac{2\pi e}{%
J(\theta )}\right\} \right\rangle _{\theta }  \label{IJ}
\end{equation}%
where $H_{\theta }=-\int d\theta p(\theta )\log _{2}p(\theta )$, in
agreement with previous results\cite{rissanen,brunel}.

We now turn to the more difficult case of discrete valued stimulus which
takes the values $\left\{ {\theta }_{l}\right\} _{l=1}^{M}$ and $M$ remains
finite as $N\rightarrow \infty $. The term $\phi =\theta $ in the average $%
\left\langle \exp S_{\phi }\right\rangle _{\phi }$ in Eq.~(\ref{DH}) is $%
p(\theta )$, yielding a total contribution to $I$ which is the stimulus
entropy $H_{\theta }=-\sum_{\theta }p(\theta )\log _{2}p(\theta )$. \
Naively, we would therefore expect that the main contribution to $\log
_{2}p(\theta )-I(\theta )$ comes from the typical values of $S_{\phi }$,
namely, $-D_{KL}(\phi ||\theta )$, for a state $\phi $ which is closest to $%
\theta .$ Such a contribution would be proportional to $\exp (-D_{KL}(\phi
||\theta ))$. \ However, we find that in fact, the dominant contribution
comes from rare values of ${\bf r}$ such that $S_{\phi }$ $\cong 0$ for one
of the states $\phi $ $(\phi \neq \theta )$. This is because, as we will
show, although this regime has an exponentially small probability, its
contribution to $I$ is exponentially larger than that of the typical value
of $S_{\phi }$ making it the dominant correction to $I-H_{\theta }$.

To evaluate Eq.~(\ref{DH}) in the discrete case we use an integral
representation of the distribution of $S_{\phi }$ $\equiv S({\bf r,}\phi
,\theta )$, 
\begin{eqnarray}
&&P_{\theta }\{S_{\phi }\}=\int \prod_{\phi \neq \theta }\frac{dY_{\phi }}{%
2\pi }\exp \{-F_{\theta }(Y_{\phi },S_{\phi })\}  \label{Integ} \\
&&F_{\theta }=-\sum_{i=1}^{N}\ln \left\langle \exp i\Sigma _{\phi }Y_{\phi
}S_{i}(r_{i},\phi ,\theta )\right\rangle _{r|\theta }+i\sum_{\phi }Y_{\phi
}S_{\phi }.  \label{F}
\end{eqnarray}%
Note that there is no integration over variables with $\phi =\theta $, since 
$S_{\theta }$ $\equiv S({\bf r,}\theta ,\theta )$ $=0$. The large $N$ limit
of Eq.~(\ref{DH}) and (\ref{Integ}) is evaluated by the saddle point method.
Solving the saddle point equations for $\{S_{\phi }\}$, and $\{Y_{\phi }\}$
we find that there are $M-1$ saddle points, each of which is characterized
by having one $S_{\phi }$ of $O(1)$ while the remaining $S_{\phi ^{\prime }}$
are negative and of $O(N)$. The auxiliary variable $Y_{\phi }=-i\alpha $
where $\alpha $ is a real number of order $1$ while the remaining $Y_{\phi
^{\prime }}$ are zero. At this saddle point, the value of $F_{\theta }$,
Eq.~(\ref{F}) is 
\begin{equation}
D_{\alpha }(\phi ||\theta )\equiv -\sum_{i=1}^{N}\ln \left\langle \exp
\alpha S_{i}(r_{i},\phi ,\theta )\right\rangle _{r|\theta }  \label{dbeta}
\end{equation}%
where the order parameter $\alpha $ is evaluated by maximizing Eq.~(\ref%
{dbeta}), 
\begin{equation}
\frac{\partial }{\partial \alpha }D_{\alpha }(\phi ||\theta )=0.
\label{beta}
\end{equation}%
We will denote the maximum value of $D_{\alpha }$ by $D_{C}(\phi ,\theta )$.
Out of these $M-1$ saddle points the dominant one is that with the smallest $%
D_{C}(\phi ,\theta )$, yielding 
\begin{equation}
\ln \left[- \log _{2}p(\theta )-I(\theta )\right] \approx -\min_{\phi \neq
\theta }D_{C}(\phi ,\theta ).  \label{dhp}
\end{equation}%
Finally, the full MI is given by 
\begin{equation}
\ln \left[ H_{\theta }-I\right] =-\min_{\left( \phi \neq \psi \right)
}D_{C}(\phi ,\psi )-\frac{1}{2}\ln N+A  \label{finalI}
\end{equation}%
where the $\ln N$ and the constant $A$ correction terms come from the
Gaussian integration around the saddle point of Eqs.~(\ref{DH}) and (\ref%
{Integ}). The calculations of the constant $A$  will be presented elsewhere %
\cite{kang}.  Finally, examining Eq.~(\ref{Integ}) at the saddle-point,
it can be seen that $%
\exp(-D_{C}(\phi ,\theta ))$ equals the probability that $S_{\phi }/N\approx 0$
for a large $N$.
This implies, as we have mentioned above, that the rate of saturation of
the MI is dominated by the probability of the rare event that one of the
log-likelihood ratios is close to zero.

The quantity $D_{C}(\phi ,\psi )$ derived in the above theory is identical to
the Chernoff Distance (or Chernoff Information) between the two
distributions $P({\bf r}|\phi )$ and $P({\bf r}|\psi )$. More generally, the
Chernoff Distance between two arbitrary distributions $P(x)$, and $\,Q(x)$,
is defined \cite{cover} as 
\begin{equation}
D_{C}(Q,P)=\max_{\alpha }D_{\alpha }(Q||P)  \label{dcgeneral}
\end{equation}%
where 
\begin{equation}
D_{\alpha }(Q||P)\equiv -\ln \sum_{x}Q(x)^{\alpha }P(x)^{1-\alpha }
\label{dqp}
\end{equation}
$D_{\alpha }(Q||P)$ are proportional to the family of Renyi distances.\cite%
{renyi}. $D_{\alpha }(Q||P)$ vanishes at $\alpha =0$ and $\alpha =1$. It is
positive (if $Q\neq P$) for $0<\alpha <1$, with a maximum at $\alpha ^{\ast }
$, $0<\alpha ^{\ast }<1$, and is negative outside this regime. It is related
to the KL distance through its slope at $\alpha =0$, {\it i.e.,} $\partial
D_{\alpha }(Q||P)/\partial \alpha |_{\alpha =0}=D_{KL}(Q||P)$. The Chernoff
Distance chooses the value of $\alpha $ which maximizes $D_{\alpha }(Q||P)
$. This value is not constant but depends on the pair of distributions $Q$
and $P$.  Note, that in the case of 
a family of distributions parameterized by a continuous
parameter $\theta $, the Chernoff Distance is related to the FI through $%
4D_{C}(\theta ,\theta +\delta \theta )\simeq D_{KL} 
(\theta, \theta +\delta\theta) \simeq J(\theta )(\delta \theta )^{2}/2$. 

Although $D_{C}(Q,P)$ is not symmetric with respect to $Q$ and $P$ for
general $\alpha $, it obeys the symmetry $D_{\alpha }(Q||P)=D_{1-\alpha
}(P||Q),$ which implies that $D_{\alpha ^{\ast }}(Q||P)=$ $D_{\alpha ^{\ast
}}(P||Q)$. Thus, the Chernoff Distance of $Q$ and $P$ is a symmetric
function of the two distributions. It is smaller than $D_{KL}$, and in
addition, it is less sensitive than the KL distance to outlier states. In
particular, a single state which has a nonzero probability $P$ but zero
probability $Q$, causes $D_{KL}(Q||P)$ to diverge whereas $D_{C}(Q,P)$
remains finite.  In fact, it diverges only when the
two distributions have zero overlap, i.e., the intersection of their support
is empty.

Equation (\ref{finalI}) implies that the Chernoff distance controls the rate
of saturation of the MI. In order to test our theory we have studied the
case of a population of statistically independent, binary neurons, where $%
r_{i}=\{0,1\},$ and a stimulus which can take three values: $\theta
_{1},\theta _{2}$ and $\theta _{3}$. Each neuron has a preferred
stimulus, denoted by $\theta ^{i}.$ 
The mean value of $r_i$ ({\it i.e.}, the probability that $r_{i}=1$)
is
\begin{equation}
\label{example1}
f_{i}(\theta)=z_{i}(\theta) +T\delta_{\theta,\theta^i}
\end{equation}
For each neuron, the $\theta _{i}$ is chosen at
random from  $\left\{ \theta _{1},\theta _{2}, \theta _{3}\right\} ,$ with
equal probability, and $z_{i}(\theta)$ 
are chosen at random uniformly
from $[a,b]$, with $T\geq b-a$.  Thus, the parameter 
$T$ measures the selectivity of the responses to the stimulus values.
In this case, there is no statistical difference between the response
probability of the population as a whole to the three stimuli, and changing
the stimulus corresponds to permuting the mean responses of the individual
neurons in the population.  This is common in biological
situations where different stimuli (such as different angles, spatial
positions or abstract objects) elicit  activity profiles that are similar in
shape but shifted in position across the network.  
\begin{figure}[tbp]
\leavevmode\centering\psfig{file=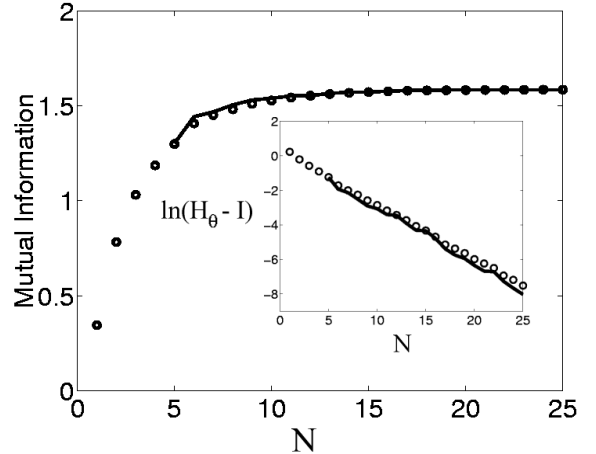,width=3.00in}
\caption{The Mutual Information between a population of $N$ independent
binary neurons and an M-state stimulus, with $M=3$, as a function of $N$. 
The response parameters $T,a,$ and $b$ are $0.75, 0.05$, and $0.15$, 
respectively. 
The dots are the exact numerical calculation. The line is 
the approximation, Eq.~(17). We have included a constant $\ln(M-1)$ to 
account for the $M-1$ equal contributions to $I(\theta)$ because of the 
symmetry of the stimuli. The Inset compares the 
two results on a log plot.}
\label{fig:binary}
\end{figure}
Figure 1 shows a nice
agreement between the results of exact numerical calculation of the MI of this 
model up to size $N=25,$ and the asymptotic result of 
Eqs.~(\ref{dbeta})-(\ref{finalI}), 
evaluated for the above distributions. Note that because of the symmetry
between the different stimuli in this example, $D_{C}(\phi ,\psi ),$ is the
same for all pairs of stimuli, $\theta \neq \theta ^{\prime }$, so that they
all contribute equally to Eq.~(\ref{finalI}).     

The importance of $D_{C}(\phi ,\psi )$ in characterizing the efficiency of
the population code is manifest not only in the MI\ of the system but also
in the accuracy of the discrimination between stimuli on the basis of
observation of the population responses. Plausible discriminatory often base
their discrimination between a pair of stimuli on the log-likelihood ratio
of the corresponding distributions. 
In particular, the Maximum-Likelihood (ML) discriminator makes a
deterministic decision between a pair of stimuli $\theta $ and $\phi $ upon
observing ${\bf r}$, according to whether the log-likelihood, $S({\bf r,}%
\phi ,\theta )$, is larger or smaller than zero. 
As outlined below, our theory yields that the probability
of confusion of an ML discrimination between a pair of stimuli,  $%
\theta $ and $\phi ,$ is dominated by the probability that $S({\bf r,}\phi
,\theta )$ is close to zero, and thus is determined by $D_C(\phi,\theta)$.
\begin{figure}[tbp]
\leavevmode\centering\psfig{file=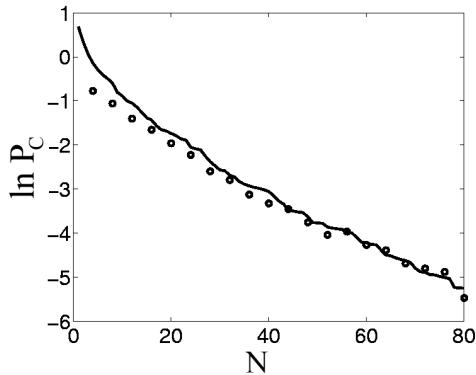,width=2.50in}
\caption{The probability of confusion in the case of an independent
population of Poissonian neurons, incurred by an ML discriminator whose task is
to discriminate between three stimuli. The mean response of each neuron is 
given by Eq. ~(20), with $T=3$, and $z_i(\theta)$ 
sampled uniformly, between $a=17$, and $b=20$. Dots are the ML 
decision error calculated by averaging over 5000 samples of $\bf{r}$.  
The line is the prediction of Eq.~(22). }
\label{fig:pc}
\end{figure}
To evaluate the discrimination error, 
we write the average probability of confusion, 
$P_C$ of an ML discrimination 
as 
$P_C= \left\langle P_C(\theta )\right\rangle _{\theta } $
where,
\begin{equation}
P_C(\theta )=1-\int \prod_{\phi \neq \theta }dS_{\phi } \Theta(-S_{\phi}) P_{\theta }\{S_{\phi
}\}\prod_{\phi\neq\theta}  \label{pctheta}
\end{equation}%
where $\Theta(x)$ is the Heaviside step function, and $P_{\theta }\{S_{\phi
}\}$ is given by Eq.~(\ref{Integ}).
Using the saddle point methods as described above,
we find that in a large system the dominant contribution to the integrals
over $S_\phi$ comes from the 
edge points where one $S_\phi$ is close to zero. Evaluating the saddle point
equations for $Y_\phi$ under this condition, yields 
\begin{equation}
\ln P_C =-\min_{\left( \phi \neq \psi \right)
}D_{C}(\phi ,\psi )-\frac{1}{2}\ln N+A'  \label{pc}
\end{equation}%
An example is shown in Fig.~2, where we have computed the confusion error of
an ML discriminator between two stimuli in the case of a population of
neurons responding to three stimuli as described above, except that in this
case, the neurons are Poissonian with means $f_{i}(\theta )$, of the form
Eq.~(\ref{example1}).
The numerical results obtained by simulating the ML discriminator
are in very good agreement with the prediction of Eq.~(\ref{pc}).

In conclusion, we have shown the relation between the MI of a large
population coding for a stimulus and the distance between the
response-probabilities that are induced by the different stimulus values. In
the case of a continuous stimulus, the MI increases logarithmically with $N$%
, for large $N$, and is related to the FI which measures the vanishing rate
of the distance for infinitesimally small stimulus differences. In the case
of discrete stimulus, the MI saturates exponentially at a rate which is
given by the Chernoff Distance between the closest pair of population
response probabilities. In addition, we have shown that
$D_C$ determines also the probability of discrimination error. This extends the 
classical Large Deviation Theory results\cite{cover}
to cases where the elements of the population are not identical.
Our finding that the Chernoff Distance controls both the MI and the error probability for 
a large population that code discrete stimuli, suggests that $D_C$ is a useful 
measure of the quality of neuronal population codes.
We hope that these
results will provide tractable tools to study the nature of population codes in
the brain using experimental data on neuronal representations of sensory,
motor and cognitive events.

This work was partially supported by grants of the Israeli Science
Foundation and the Israel-USA Binational Science Foundation. We thank
Samengo and Treves for discussions on Ref.\cite{samengo}.

\end{multicols}

\end{document}